\documentclass[10pt,a4paper]{article}
\usepackage[utf8]{inputenc}
\usepackage[english]{babel}
\usepackage[OT1]{fontenc}
\usepackage{tikz}
\usepackage{amsmath} 
\usepackage{amsfonts}
\usepackage{amssymb}
\usepackage{amsthm}
\usepackage{amscd}
\usepackage{amsfonts}
\usepackage{graphicx}
\usepackage{fancyhdr}
\usepackage{pbox}

\providecommand{\pacs}[1]{PACS numbers : #1}
\providecommand{\keywords}[1]{Keywords : #1}

\usepackage{longtable}
\usepackage{hyperref}
\usepackage{lipsum}

\usepackage[margin=1.25in,footskip=0.25in]{geometry}

\begin{document}

\title{Maxwell's equations and Lorentz force in doubly special relativity}
\author{N. Takka\hspace{0.1cm}\footnote{E-mail: takka.naimi@gmail.com} \hspace{0.1cm}and\hspace{0.1cm} A. Bouda\hspace{0.1cm}\footnote{E-mail: bouda-a@yahoo.fr}\hspace{0.1cm}\\  Laboratoire de Physique Théorique, Faculté des Sciences Exactes,\\
Université de Bejaia, 06000 Bejaïa, Algeria}
\date{\today}
\maketitle

\begin{abstract}
 On the basis of all commutation relations of the $\kappa$-deformed phase space incorporating the $\kappa$-Minkowski space-time, we have derived in this paper an extended first approximation of both  Maxwell's equations and Lorentz force in doubly (or deformed) special relativity (DSR). For this purpose, we have used our approach of the special relativistic version of Feynman's proof by which we have established the explicit formulations of electric and magnetic fields. As in Fock's nonlinear relativity (FNLR), the laws of electrodynamics depend on the particle mass which therefore constitutes a common point between the two extended forms of special relativity. As one consequence, the corresponding equation of motion contains two different types of contributions. In addition to the usual type, another one emerges as a consequence of the coexistence of mass and charge which are coupled with the $\kappa$-deformation and electromagnetic field. This new effect completely induced by the $\kappa$-deformed phase space is interpreted as the gravitational-type Lorentz force. Unlike FNLR, the corrective terms all depend on the electromagnetic field in DSR.
\end{abstract}
       
\keywords{Noncommutative geometry, $\kappa$-Minkowski space-time, Maxwell's equations}

\pacs{02.40.Gh, 03.30.+p, 03.50.De} 

\newpage

 
 

\section{Introduction}
In order to reconcile general relativity and quantum mechanics, several avenues of research have been proposed. Among the most developed theories in the last decades, it is well-known that string theory and loop quantum gravity continue to attract a lot of attention and might constitute viable candidates for a unified theory. So far, due to the lack of testable predictions to rule in favor or disfavor of either (or neither) of these two theories, on the one hand, curiosity driven by the current theoretical and practical success, on the other hand, new ideas appear as a source of motivation encouraging the emergence of new theories. In what concerns us here, we hint at doubly special relativity also known as deformed special relativity or simply DSR \cite{G.Amelino-Camelia-1}-\cite{Ghosh-Pal-1}. The main novelty of such extension is the possibility to relate the two pillars of modern physics where the boundary is set by Planck scale, or more precisely, Planck energy $E_{p}$. In this perspective, DSR represents an extended formulation of special relativity where Planck energy $E_{p}$, or its inverse, Planck length ($l_{p}=\hbar c/E_{p}=\hbar/\kappa$) joins the speed of light as a second invariant.

In the quest for new laws of electromagnetism even more general, many attempts are developed in the framework of commutative and noncommutative spaces. One of mathematical tools devoted to this goal is known as Feynman's proof \cite{Dyson}. In its original form, the usual homogeneous Maxwell's equations and Lorentz force are reproduced by assuming Newton’s equation of motion and the commutation relation between position and velocity for a non-relativistic particle. To be in line with some different areas of mathematical physics research, this approach has undergone several developments. The first significant advance could be associated to the formulation of both special and general relativistic versions of this derivation \cite{S.Tanimura-1} as a response to some reactions of the scientific community \cite{Dombey-1}-\cite{Vaidya-Farina-1}. In this work, it is  shown that the only possible fields that can consistently act on a quantum particle are scalar, gauge and gravitational fields. In \cite{Hojman-Shepley-1} and \cite{Hughes-1},  some interesting aspects of comparability with the Lagrangian description are raised. Subsequently, the use of Hodge duality made possible the derivation of the two groups of Maxwell's equations with a magnetic monopole in flat and curved spaces \cite{Berard-Grandati-Mohrbach-1}. In other studies, the investigation of Dirac magnetic monopole gave rise to some other discussions, rederivations and applications \cite{Berard-Grandati-Mohrbach-2}, \cite{Berard-Mohrbach-1} and \cite{Berard-Lages-Mohrbach-1}. In \cite{Montesinos-Perez-Lorenzana-1}, a new look on Feynman's proof is provided and its equivalence with minimal coupling is discussed. Due to the increasing interest in noncommutative formalism, the generalization of Feynman's approach and its various physical and mathematical interpretations are gaining a lot of attention in this specific area, e.g. \cite{Boulahoual-Sedra-1}, \cite{Carinena-Figueroa-1}, \cite{Berard-Mohrbach-Lages-Grandati-Boumrar-Boumrar-Menas-1}. As a relative application, an insight for the study of the electric-magnetic duality is highlighted in \cite{Boulahoual-Sedra-1}, the impact of noncommutativity on the particle dynamics is evaluated in \cite{Carinena-Figueroa-1} and a new effect called the spin Hall effect is revealed in \cite{Berard-Mohrbach-Lages-Grandati-Boumrar-Boumrar-Menas-1}. In a related context, a variational principle is suggested for noncommutative dynamical systems in configuration space \cite{Cortese-Garcia-2}. Moving forward to $\kappa$-Minkowski space-time, it turned out that the first-order approximation of the laws of electrodynamics depends on the particle mass \cite{Harikumar-1} and \cite{Harikumar-Juric-Meljanac-1}. In \cite{Harikumar-Juric-Meljanac-1}, this is done by power series expansion of noncommutative coordinates, momenta and some functions depending on them, in terms of commutative coordinates, momenta and deformation parameter. After that, the linear approximation of the $\kappa$-deformed geodesic equation is derived and some interpretations are given in \cite{Harikumar-Juric-Meljanac-2}. Recently, we have succeeded to go beyond our first-order generalization \cite{Takka-Bouda-Foughali-1}, by establishing the exact formulation of  Maxwell's equations, Dirac's magnetic monopole \cite{Takka-Bouda-2} and Lorentz force \cite{Takka-3} in the context of Fock-Lorentz transformation \cite{V.Fock-1}-\cite{Bouda-Foughali-1}. By restoring the deformed Lorentz algebra, we have predicted that the universe could locally contain the magnetic charge but in its totality it is still neutral \cite{Takka-Bouda-2}. In \cite{Takka-3}, the generalized Lorentz force is characterized by a separability between mass and charge where the new effect emerges as the result of coexistence of the radius of the universe, mass and the square of the velocity, but completely independent of
 electromagnetic field (gravitational-type Lorentz force).

This paper is organized as follows. In the next section, we have established an extended first approximation of Maxwell's equations in DSR. To this end, we have used our approach of the special relativistic version of Feynman's proof where all commutation relations of the $\kappa$-deformed phase space are taken into account. At the end, we have also exploited the final result of Appendix in order to obtain the corresponding explicit forms of both electric and magnetic fields. In the same way, we have derived in section $3$ the generalized Lorentz force and shown that this latter partially depends on the particle mass. The results and discussion are presented in section $4$. Section $5$ is devoted to conclusion.

 
 

\section{$\kappa$-Generalized Maxwell's equations}
By modifying the usual Poisson brackets, they suggested in \cite{Ghosh-Pal-1} a more familiar  version of DSR, previously constructed in \cite{G.Amelino-Camelia-1}-\cite{Magueijo-Smolin-2} whose first writing of the $\kappa$-deformed phase space dates back to  \cite{Lukierski-Nowicki-Ruegg-1}- \cite{Lukierski-1}. After a first quantization, this new version is characterized by the following commutation relations

\begin{align}
\left[x^\mu,x^\nu\right] & = \frac{i\hbar}{\kappa}\left(x^{\mu}\eta^{\nu 0}-x^{\nu}\eta^{\mu 0}\right),\label{equation-1}\\
\left[x^\mu,p^\nu\right] & = i\hbar\big(-\eta^{\mu\nu}+\frac{1}{\kappa}\eta^{\mu 0}p^{\nu}\big), \label{equation-2}\\
\left[p^\mu,p^\nu\right] & = 0, \label{equation-3}
\end{align}

\noindent where $\eta^{\mu\nu}=(1,-1,-1,-1)$, $\mu ,\nu =0,1,2,3$ and $\kappa=\left(E_{p}/c\right)$ is identified as a deformed parameter, $E_{p}$ and $c$ represent Planck energy and the speed of light in a vacuum respectively.  Obviously, if we take the simple limit $\hbar\rightarrow 0$, the noncommutative space-time considered above does not lead to the usual $\kappa$-Minkowski space-time (the Poisson brackets are replaced by commutators in the following rule $\{ \ \}\mapsto \frac{1}{i\hbar}[ \ ]$).  In \cite{Ghosh-Pal-1}, the $\kappa$-deformed phase space allows the reproduction of well-known momentum and coordinate transformations

\begin{equation}\label{equation-4}
E^{\prime }=\frac{\gamma\left(E-up_{x}\right)}{\alpha_\kappa},\quad p_{x}^{\prime }=\frac{\gamma\left(p_x-uE/c^{2}\right)}{\alpha_\kappa},
\quad p_y^{\prime }=\frac{p_y}{\alpha_\kappa},\quad p_z^{\prime}=\frac{p_z}{\alpha_\kappa}
\end{equation}                                          

\noindent and                                                                         

\begin{equation}\label{equation-5}
t^{\prime }=\alpha_{\kappa}\gamma\left(t-ux/c^2\right),\quad x^{\prime}=\alpha_{\kappa}\gamma\left(x-ut\right),\\ \quad y^{\prime }=\alpha_\kappa y,\quad z^{\prime}=\alpha_\kappa z.
\end{equation}

\noindent Here $u$ represents the relative velocity between two inertial frames in the $x$-direction, $\gamma=\left(1-u^{2}/c^{2}\right)^{-1/2}$ and $\alpha_\kappa=1+\left[\left(\gamma-1\right)E-\gamma u p_{x}\right]/E_{p}$. It is possible to check that Planck energy $E_{p}$, or its inverse, Planck length ($l_{p}=\hbar c/E_{p}=\hbar/\kappa$) is kept invariant due to the fact that $E^{\prime }_{p}=E_{p}$. We can easily check that in the limit where $\kappa\rightarrow\infty$, we recover all conventional results. Up to first order in the deformation parameter $\kappa$, we can write 

\begin{align}
x^{\mu}_{(1)} & = x^{\mu} + \delta x^{\mu},\label{equation-6}\\
p^{\mu}_{(1)} & = p^{\mu} + \delta p^{\mu}. \label{equation-7}
\end{align}

\noindent Obviously $\delta x^{\mu}$ and $\delta p^{\mu}$ are both  functions of position and momentum which are linear in $\kappa^{-1}$. To seek the appropriate corrective terms for both position and momentum operators, we rely on Eqs. (\ref{equation-1}), (\ref{equation-2}) and (\ref{equation-3}) as constraints. Indeed, the use of Eq. (\ref{equation-6}) gives 

\begin{eqnarray}\label{equation-8}
\left[x^{\mu} , x^{\nu} \right]_{(1)} &=&\left[x^{\mu}, p^{\alpha}\right]_{(0)}\frac{\partial \delta x^{\nu}}{\partial p^{\alpha}}+\left[p^{\alpha}, x^{\nu} \right]_{(0)}\frac{\partial \delta x^{\mu}}{\partial p^{\alpha}}\nonumber\\
       &=& i\hbar\left(\frac{\partial \delta x^{\mu}}{\partial p_{\nu}}-\frac{\partial \delta x^{\nu}}{\partial p_{\mu}}\right),
\end{eqnarray}

\noindent which, after identification with Eq. (\ref{equation-1}), yields

\begin{equation}\label{equation-9}
\frac{\partial \delta x^{\mu}}{\partial p_{\nu}}-\frac{\partial \delta x^{\nu}}{\partial p_{\mu}} = \frac{1}{\kappa}\left( x^{\mu}\eta^{\nu 0}-x^{\nu}\eta^{\mu 0}\right).
\end{equation}

\noindent In the same way, the use of Eqs. (\ref{equation-6}) and (\ref{equation-7}) into (\ref{equation-2}) and (\ref{equation-3}) implies respectively 

\begin{align}
\frac{\partial \delta p^{\nu}}{\partial p_{\mu}}+\frac{\partial \delta x^{\mu}}{\partial x_{\nu}} & = - \frac{1}{\kappa}\eta^{\mu 0}p^{\nu},\label{equation-10}\\
\frac{\partial \delta p^{\nu}}{\partial x_{\mu}}-\frac{\partial \delta p^{\mu}}{\partial x_{\nu}} & = 0. \label{equation-11}
\end{align}

\noindent Mathematically, we can make sure that Eqs. (\ref{equation-9}) and (\ref{equation-11}) are satisfied respectively by the following general solutions

\begin{align}
 \delta x^{\mu} & = \frac{\alpha}{\kappa}\left<p^{0} x^{\mu}\right> +  \frac{\beta}{\kappa} \left<x^{0} p^{\mu}\right> + \frac{\gamma}{\kappa} \eta^{\mu 0}\left<x_{\lambda} p^{\lambda}\right>,\label{SG-dx}\\
 \delta p^{\mu} & =  \frac{\zeta}{\kappa} p^{0} p^{\mu} + \frac{\xi}{\kappa}\eta^{\mu 0} p_{\lambda} p^{\lambda} + 
 \frac{\chi}{\kappa} \left<p^{0} x^{\mu}\right>\label{SG-dp},
 \end{align}

\noindent where $\alpha, \beta, \gamma, \zeta \hspace*{0.1cm}\text{and}\hspace*{0.1cm}\xi \in\mathbb{R^{*}}$, $\chi$ is a dimensional constant and $<.>$ refers to the symmetrization operator \cite{S.Tanimura-1}. Furthermore, if we substitute the two last results into (\ref{equation-10}), we can easily see that

\begin{equation}\label{zeta-gamma-alpha-xi-beta-chi-1}
 \zeta + \gamma = -1,\hspace*{0.2cm}\zeta + \alpha = 0,\hspace*{0.2cm}2\xi + \beta = 0,\hspace*{0.2cm}\chi = 0.
\end{equation}

\noindent The above result can be related to some realizations of $\kappa$-Minkowski space [\cite{Harikumar-Juric-Meljanac-1}, \cite{Juric-Meljanac-Pikutic-1}, \cite{Kovacevic-Meljanac-Samsarov-Skoda-1}] and the study focused on $SU(2)$ case \cite{Juric-Poulain-Wallet-1}. To further reduce the number of parameters appearing above, let us recall that Ghosh and Pal have reproduced in \cite{Ghosh-Pal-1} the well-known Magueijo-Smolin dispersion relation \cite{Magueijo-Smolin-1}

\begin{equation}\label{MSDR}
(1-\frac{E}{E_{p}})^{-2}\left(E^{2}-\vec{p}^{2}c^{2}\right)=m^{2}_{0}c^{4}.
\end{equation}

\noindent Before continuing, let us note that in an interesting work, a general framework describing modified dispersion relations and time delay with respect to different noncommutative $\kappa$-Minkowski spacetime realizations is firstly proposed in \cite{Borowiec-Gupta-Meljanac-1} and it covers all the cases introduced in the literature. By analyzing the noncommutative version of the free classical field theory, the impact of the noncommutativity on the dispersion relations is shown in \cite{Juric-Meljanac-Pikutic-Strajn-1}. Recently, by considering the $\kappa$-deformed relativistic quantum phase space and possible implementations of Lorentz algebra, the dispersion relations and their physical consequences are discussed in \cite{Meljanac-Meljanac-Mignemi-Strajn-1}. For a massless particle, the last result reduces to the usual form known in special relativity. Now, to be able to use (\ref{MSDR}) in order to find the four-momentum which is compatible with our context, let us begin by the generalization of  (\ref{equation-4}) to the case where the translational motion between the frames $(R)$ and $(R^{\prime})$ is not necessarily collinear to the $x$-axis $(\vec{p}=\vec{p}_{\parallel} + \vec{p}_{\perp})$. In this case, the corresponding inverse transformation with respect to the velocity $(\vec{u}\rightarrow -\vec{u})$ takes this form

\begin{equation}\label{GKLT}
E= \frac{\gamma}{\alpha^{\prime}_{\kappa}}\left(E^{\prime }+\vec{u}\cdot\vec{p}^{\prime}_{\parallel}\right), \quad \vec{p}_{\parallel}= \frac{\gamma}{\alpha^{\prime}_{\kappa}}\left(\vec{p}^{\prime}_{\parallel}+\frac{\vec{u}}{c^{2}} E^{\prime}\right),\quad \vec{p}_{\perp}= \frac{\vec{p}^{\prime}_{\perp}}{\alpha^{\prime}_{\kappa}},
\end{equation} 

\noindent  where $\alpha^{\prime}_{\kappa}=1+\left[\left(\gamma-1\right)E^{\prime}+\gamma \vec{u}\cdot\vec{p}^{\prime}_{\parallel}\right]/E_{p}$. To express the energy and momentum of our $\kappa$-particle with respect to its three-dimensional velocity $v^{i}$, we introduce a third frame $(R_{0})$ which is attached to this latter and then relatively to $(R)$, we get $(u^{i}=v^{i})$. Since the transition from $(R_{0})$ to $(R)$ can be considered locally inertial, the above relations become

\begin{equation}\label{GKLT-PC}
E= \frac{\gamma_{0}}{\alpha^{0}_{\kappa}}E_{0}, \quad \vec{p}_{\parallel}= \frac{\gamma_{0}}{\alpha^{0}_{\kappa}}E_{0}\left( \frac{\vec{v}}{c^{2}} \right),\quad \vec{p}_{\perp}= \vec{0},
\end{equation}

\noindent  here $\alpha^{0}_{\kappa}=1+\left(\gamma_{0}-1\right)E_{0}/E_{p}$ and $\gamma_{0}=\left(1-v^{2}/c^{2}\right)^{-1/2}$. To move forward, let us recall that the modified Lorentz transformations give the following invariant \cite{Ghosh-Pal-1}

\begin{equation}\label{Invariant-DSR-P}
I_{p}=(1-\frac{p_{t}'}{k})^{-2}p'_{\mu}p'^{\mu}=(1-\frac{p_{t}}{k})^{-2}p_{\mu}p^{\mu},
\end{equation}

\noindent which, relatively to the frame at rest, yields

\begin{equation}\label{Invariant-DSR-E_0}
(1-\frac{E}{E_{p}})^{-2}\left(E^{2}-\vec{p}^{2}c^{2}\right)=(1-\frac{E_{0}}{E_{p}})^{-2} E_{0}^{2}.
\end{equation}

\noindent  At this stage of development, a simple identification between the last result and Eq. (\ref{MSDR}) gives the relationship between energy and mass at rest

\begin{equation}\label{DSR-E_0-m_0}
E_{0}=\frac{m_{0}c^{2}}{1+\frac{m_{0}c^{2}}{E_{P}}},
\end{equation}

\noindent which, after substitution into $\alpha^{0}_{\kappa}$, yields

\begin{equation}\label{alpha-0_k}
\alpha^{0}_{\kappa}=\frac{E_{P}+\gamma_{0}m_{0}c^{2}}{E_{P}+m_{0}c^{2}}.
\end{equation}

\noindent Finally, after substitution of (\ref{DSR-E_0-m_0}) and (\ref{alpha-0_k}) into (\ref{GKLT-PC}), we reproduce the four-momentum  previously found  by Magueijo-Smolin \cite{Magueijo-Smolin-1}

\begin{align}
p^{0} &= \frac{mc}{1+\frac{mc}{\kappa}},\\
p^{i} &= \frac{m v^{i}}{1+\frac{mc}{\kappa}}.
\end{align}

\noindent Above $m=\gamma_{0}m_{0}$. In the presence of electromagnetic field and at first order in $1/\kappa$, the covariant formulation of the two last relations takes the following form

\begin{equation}\label{equation-12}
p^{\mu}_{(1)} = \left(1-\frac{p^{0}}{\kappa}\right)p^{\mu}. 
\end{equation}

\noindent The above result remains unchanged in the quantum regime since $\left[p^\mu,p^\nu\right]=0$. By using the canonical variables found in \cite{Ghosh-Pal-1}, we can easily check the correctness of the last result. Taking into account (\ref{equation-12}), Eqs. (\ref{SG-dp}) and (\ref{zeta-gamma-alpha-xi-beta-chi-1}) lead to

\begin{equation}\label{zeta-beta-gamma-xi-alpha-chi-2}
\zeta=-1, \hspace*{0.2cm} \beta=\gamma=\xi=\chi=0,\hspace*{0.2cm} \alpha=1,
\end{equation}

\noindent which, after substitution into (\ref{SG-dx}), yields

\begin{equation} \label{equation-13}\\
 \delta x^{\mu}  = \frac{1}{\kappa}\left<p^{0} x^{\mu}\right>=\frac{1}{2 \kappa}\left(p^{0} x^{\mu} + x^{\mu}p^{0}\right)=\frac{1}{\kappa} p^{0} x^{\mu}-\frac{i\hbar}{2\kappa}\eta^{\mu 0}
\end{equation}

\noindent and then Eq. (\ref{equation-6}) becomes

\begin{equation}\label{equation-14}
x^{\mu}_{(1)} = \left(1+\frac{p^{0}}{\kappa}\right)x^{\mu}-\frac{i\hbar}{2\kappa}\eta^{\mu 0}.
\end{equation}

\noindent As it has already been mentioned above for (\ref{equation-12}), the canonical variables defined in \cite{Ghosh-Pal-1} lead to the last result after symmetrization. Now, to derive the generalized Maxwell's equations valid up to first order in $1/\kappa$,  we proceed as in \cite{S.Tanimura-1} by starting from the following Jacobi identity

\begin{equation}\label{equation-16}
\left[\dot{x}^{\mu}, \left[\dot{x}^{\nu}, \dot{x}^{\lambda}\right]\right]_{(1)} + \left[\dot{x}^{\lambda}, \left[\dot{x}^{\mu} , \dot{x}^{\nu}\right]\right]_{(1)} + \left[\dot{x}^{\nu} , \left[\dot{x}^{\lambda} , \dot{x}^{\mu}\right]\right]_{(1)} = 0.
\end{equation}

\noindent Without doing any calculation, from (\ref{equation-3}) and (\ref{equation-12}) it is predictable that the generalized electromagnetic tensor depends on both position and velocity. Therefore, to the first order in $1/\kappa$, $F^{\mu\nu}_{(1)}= F^{\mu\nu}(x) + \delta F^{\mu\nu}(x,\dot{x})$ and can be introduced as follows

\begin{equation}\label{equation-17}
\left[\dot{x}^{\mu}, \dot{x}^{\nu}\right]_{(1)}=-\frac{i\hbar q}{m^{2}}F^{\mu\nu}_{(1)}(x,\dot{x}).
\end{equation}

\noindent As $\delta F^{\mu\nu}(x,\dot{x})$ does not contain quadratic terms in velocity, it is obvious that 

\begin{equation}\label{equation-18}
\frac{\partial F^{\mu\nu}_{(1)}}{\partial \dot{x}_{\lambda}}=\frac{\partial \delta F^{\mu\nu}}{\partial \dot{x}_{\lambda}}(x),
\end{equation}

\noindent which, by combining it with (\ref{equation-17}), gives

\begin{eqnarray}\label{equation-19}
\left[\dot{x}^{\mu}, \left[\dot{x}^{\nu}, \dot{x}^{\lambda}\right]\right]_{(1)} &=& -\frac{i\hbar q}{m^{2}}\left[\dot{x}^{\mu} + \delta \dot{x}^{\mu}, F^{\nu\lambda}(x) + \delta F^{\nu\lambda}(x,\dot{x})\right]_{(1)}\nonumber\\
&=& -\frac{i\hbar q}{m^{2}}\left\{\left[\dot{x}^{\mu}, F^{\nu\lambda}(x)\right]_{(0)} + \left[\dot{x}^{\mu}, \delta F^{\nu\lambda}\right]_{(0)}+\left[\delta\dot{x}^{\mu}, F^{\nu\lambda}(x)\right]_{(0)} \right\}\nonumber\\
&=&\frac{\hbar^{2} q}{m^{3}}\left\{\partial^{\mu} F^{\nu\lambda}_{(1)}-\frac{q}{m}F^{\mu\beta}\frac{\partial F^{\nu\lambda}_{(1)}}{\partial \dot{x}^{\beta}}\right\}-\frac{i\hbar q}{m^{2}}\left[\delta\dot{x}^{\mu}, F^{\nu\lambda}(x)\right]_{(0)}. 
\end{eqnarray}

\noindent To further develop Eq. (\ref{equation-19}), we need to know the explicit form of the remaining unknown last commutator. To this end, differentiating Eq. (\ref{equation-13}) with respect to the time parameter $\tau$ \cite{S.Tanimura-1}, to get

\begin{equation}\label{equation-20}
\delta \dot{x}^{\mu} = \frac{1}{\kappa}\left(\left<\dot{p}^{0}x^{\mu}\right> + \left<p^{0}\dot{x}^{\mu}\right>\right)
                     = \frac{1}{2 \kappa}\left(\dot{p}^{0}x^{\mu} + x^{\mu}\dot{p}^{0}\right) + \frac{1}{2\kappa}\left(p^{0}\dot{x}^{\mu}+\dot{x}^{\mu}p^{0}\right).
\end{equation}

\noindent Taking into account (\ref{equation-20}) and using the fact that in special relativity \cite{S.Tanimura-1}

\begin{align}
p^{\mu} & = m\dot{x}^{\mu} + qA^{\mu}(x),\quad c \equiv 1, \label{equation-21}\\
F^{\mu} & = m\ddot{x}^{\mu} = q\left <F^{\mu\beta}(x)\dot{x}_{\beta}\right> + G^{\mu}(x),  \label{equation-22}
\end{align}

\noindent where $F^{\mu\nu}(x) = \partial^{\mu}A^{\nu}(x) - \partial^{\nu}A^{\mu}(x)$ and $G^{\mu}(x)=\partial^{\mu}\phi$, it follows that the rewriting  where $\dot{A}^{0}(x)=\left<\partial_{\alpha}A^{0}(x)\dot{x}^{\alpha}\right>$ yields

\begin{eqnarray}\label{equation-23}
[\delta\dot{x}^{\mu}, F^{\nu\lambda}(x)]_{(0)} &=& \frac{1}{\kappa}\left(qF^{0\beta}(x) x^{\mu} [\dot{x}_{\beta}, F^{\nu\lambda}(x)]_{(0)} + m\left< \dot{x}^{0}[\dot{x}^{\mu}, F^{\nu\lambda}(x)]_{(0)} + \dot{x}^{\mu}[\dot{x}^{0}, F^{\nu\lambda}(x)]_{(0)}\right>\right.\nonumber\\
                                               &+& \left. q\partial_{\beta}A^{0}(x)x^{\mu} [\dot{x}^{\beta}, F^{\nu\lambda}(x)]_{(0)}+q A^{0}(x) [\dot{x}^{\mu}, F^{\nu\lambda}(x)]_{(0)}\right)\nonumber\\
                                               &=&\frac{i\hbar q}{m\kappa}\left(\partial^{0}A^{\beta}(x)\partial_{\beta}F^{\nu\lambda}(x)x^{\mu}+A^{0}(x)\partial^{\mu}F^{\nu\lambda}(x)\right)\nonumber\\
                                               &+& \frac{i\hbar}{\kappa}\left(\left<\dot{x}^{0}\partial^{\mu}F^{\nu\lambda}(x)\right> + \left<\dot{x}^{\mu}\partial^{0}F^{\nu\lambda}(x)\right>\right).
\end{eqnarray}

\noindent  After substitution of the last result into (\ref{equation-19}), we find

\begin{eqnarray}\label{equation-24}
\left[\dot{x}^{\mu}, \left[\dot{x}^{\nu}, \dot{x}^{\lambda}\right]\right]_{(1)} 
&=&\frac{\hbar^{2} q}{m^{3}}\left\{\partial^{\mu} F^{\nu\lambda}_{(1)}(x,\dot{x})-\frac{q}{m}F^{\mu\beta}(x)\frac{\partial F^{\nu\lambda}_{(1)}(x,\dot{x})}{\partial \dot{x}^{\beta}} \right.\nonumber\\
&+&\left.\frac{q}{\kappa}\left(\partial^{0}A^{\beta}(x)\partial_{\beta}F^{\nu\lambda}(x)x^{\mu}+A^{0}(x)\partial^{\mu}F^{\nu\lambda}(x)\right)\right.\nonumber\\
&+&\left.\frac{m}{\kappa}\left(\left<\dot{x}^{0}\partial^{\mu}F^{\nu\lambda}(x)\right> + \left<\dot{x}^{\mu}\partial^{0}F^{\nu\lambda}(x)\right>\right)\right\}.
\end{eqnarray}

\noindent Making use of Eq. (\ref{equation-24}), the development of (\ref{equation-16}) gives rise to the following first approximation of the homogeneous Maxwell's equations 

\begin{eqnarray}\label{equation-25}
&&\partial^{\mu} F^{\nu\lambda}_{(1)}+\partial^{\lambda} F^{\mu\nu}_{(1)}+\partial^{\nu} F^{\lambda\mu}_{(1)} = \frac{q}{m}\left(F^{\mu\beta}_{(0)}\frac{\partial F^{\nu\lambda}_{(1)}}{\partial \dot{x}^{\beta}} + F^{\lambda\beta}_{(0)}\frac{\partial F^{\mu\nu}_{(1)}}{\partial \dot{x}^{\beta}} + F^{\nu\beta}_{(0)}\frac{\partial F^{\lambda\mu}_{(1)}}{\partial \dot{x}^{\beta}}\right)\nonumber\\
&&  -\frac{q}{\kappa}\left(\partial^{0}A^{\beta}\right)\left(\partial_{\beta}F^{\nu\lambda}_{(0)} x^{\mu} + \partial_{\beta}F^{\mu\nu}_{(0)} x^{\lambda} + \partial_{\beta}F^{\lambda\mu}_{(0)} x^{\nu}\right) - \frac{m}{\kappa} \Big(\big<\dot{x}^{\mu}\partial^{0}F^{\nu\lambda}_{(0)}\big> \nonumber\\
&& + \big<\dot{x}^{\lambda}\partial^{0}F^{\mu\nu}_{(0)}\big> + \big<\dot{x}^{\nu}\partial^{0}F^{\lambda\mu}_{(0)}\big>\Big).
\end{eqnarray}

\noindent  To seek the corresponding second group, we proceed as in \cite{Berard-Grandati-Mohrbach-1} and \cite{Harikumar-Juric-Meljanac-1} where the starting point is the following identity

\begin{equation}\label{equation-26}
\left[\dot{x}_{\nu}, \left[\dot{x}_{\mu}, \left[\dot{x}^{\mu}, \dot{x}^{\nu}\right]\right]\right]=0.
\end{equation}

\noindent According to (\ref{equation-17}) and (\ref{equation-24}), the four-current density $J^{\nu}$ can be defined as

\begin{equation}\label{equation-27}
 \left[\dot{x}_{\mu}, \left[\dot{x}^{\mu}, \dot{x}^{\nu}\right]\right]=\frac{\hbar^{2} q}{m^{3}}J^{\nu},
\end{equation}

\noindent which, after insertion into (\ref{equation-26}), gives rise to the following conservation law 

\begin{equation}\label{equation-28}
 \left[\dot{x}_{\mu}, J^{\mu}\right]=0.
\end{equation}

\noindent  By means of Eqs. (\ref{equation-24}) and (\ref{equation-27}), we can see that the generalized inhomogeneous Maxwell's equations take this condensed form

\begin{eqnarray}\label{equation-29}
&&\partial_{\mu} F^{\mu\nu}_{(1)}= J^{\nu} + \frac{q}{m} F_{\mu\beta}(x)\frac{\partial F^{\mu\nu}_{(1)}}{\partial \dot{x}_{\beta}} -\frac{q}{\kappa}\left(\partial^{0}A^{\beta} \partial_{\beta}F^{\mu\nu}_{(0)} x_{\mu} + A^{0}\partial_{\mu}F^{\mu\nu}_{(0)}\right)\nonumber\\ 
&& - \frac{m}{\kappa} \Big(\big<\dot{x}^{0}\partial_{\mu}F^{\mu\nu}_{(0)}\big> +\big<\dot{x}_{\mu}\partial^{0}F^{\mu\nu}_{(0)}\big>\Big).
\end{eqnarray}

\noindent Finally, to get a better idea about the corrective terms appearing above, we explicitly
write (\ref{equation-25}) and (\ref{equation-29}) so that the generalized Maxwell's equations become

\begin{eqnarray}
&&\hspace*{-0.4cm} \overrightarrow{\nabla}\cdot\overrightarrow{B}_{(1)} = - \frac{q}{m}\left\{\overrightarrow{E}_{(0)} \partial_{\dot{x}^{0}} \overrightarrow{B}_{(1)} + \overrightarrow{B}_{(0)}\left(\overrightarrow{\nabla}_{v}\wedge\overrightarrow{B}_{(1)}\right)\right\} + \frac{q}{\kappa}\left(\partial_{t}A^{\mu} \partial_{\mu}\overrightarrow{B}_{(0)}\cdot\overrightarrow{r}\right)\nonumber\\ 
&&\hspace*{1.7cm}+\frac{m}{\kappa} \Big(\big<\overrightarrow{v}\cdot\partial_{t}\overrightarrow{B}_{(0)}\big>\Big),\label{equation-30}\\
&&\hspace*{-0.4cm} \overrightarrow{\nabla} \wedge \overrightarrow{E}_{(1)} + \partial_{t} \overrightarrow{B}_{(1)} = - \frac{q}{m} \left\{\left(\overrightarrow{E}_{(0)} \wedge \partial_{\dot{x}^{0}} \overrightarrow{E}_{(1)} \right) - \overrightarrow{B}_{(0)}\left(\overrightarrow{\nabla}_{v}\cdot \overrightarrow{E}_{(1)}\right) + \left(B^{i}_{(0)}\overrightarrow{\nabla}_{v}E^{i}_{(1)}\right)\right\}\nonumber\\
&&\hspace*{3.6cm} -\frac{q}{\kappa}\left\{\partial_{t}A^{\mu}\left(\partial_{\mu} \overrightarrow{E}_{(0)}\wedge \overrightarrow{r} + \partial_{\mu}\overrightarrow{B}_{(0)}x^{0}\right)\right\}\nonumber\\
&&\hspace*{3.6cm}  + \frac{m}{\kappa}\left(\left<\overrightarrow{v}\wedge \partial_{t}\overrightarrow{E}_{(0)}\right>-\left<\dot{x}^{0}\partial_{t}\overrightarrow{B}_{(0)} \right>\right),\label{equation-31}\\
&&\hspace*{-0.4cm} \overrightarrow{\nabla}\cdot\overrightarrow{E}_{(1)} = J^{0} - \frac{q}{m}\left\{\overrightarrow{E}_{(0)} \partial_{\dot{x}^{0}} \overrightarrow{E}_{(1)} + \left(\overrightarrow{B}_{(0)}\wedge\overrightarrow{\nabla}_{v}\right)\overrightarrow{E}_{(1)}\right\} - \frac{q}{\kappa} \left(A^{0}\overrightarrow{\nabla}\cdot\overrightarrow{E}_{(0)} \right.\nonumber\\ 
&&\hspace*{2.2cm}\left. -\partial_{t}A^{\mu}\partial_{\mu}\overrightarrow{E}_{(0)}\cdot\overrightarrow{r} \right)-\frac{m}{\kappa} \Big(\big<\dot{x}^{0}\overrightarrow{\nabla}\cdot\overrightarrow{E}_{(0)}\big> - \big<\overrightarrow{v}\cdot\partial_{t}\overrightarrow{E}_{(0)}\big>\Big),\label{equation-32}\\
&&\hspace*{-0.4cm} \overrightarrow{\nabla} \wedge \overrightarrow{B}_{(1)} - \partial_{t} \overrightarrow{E}_{(1)} = \overrightarrow{J} + \frac{q}{m} \left\{\left(E^{i}_{(0)}\partial_{\dot{x}^{i}} \overrightarrow{E}_{(1)} \right)-\left(\overrightarrow{E}_{(0)} \wedge \partial_{\dot{x}^{0}} \overrightarrow{B}_{(1)} \right)-\left(B^{i}_{(0)}\overrightarrow{\nabla}_{v}B^{i}_{(1)}\right)\right. \nonumber\\
&& \left. \hspace*{4cm} + \overrightarrow {B}_{(0)} \left(\overrightarrow{\nabla}_{v}\cdot\overrightarrow {B}_{(1)} \right) \right\} 
+\frac{q}{\kappa}\left\{\partial_{t}A^{\mu}\left(\partial_{\mu} \overrightarrow{E}_{(0)}x_{0} -\partial_{\mu}\overrightarrow{B}_{(0)} \wedge \overrightarrow{r}\right)\right.\nonumber\\
&&\left.  \hspace*{4cm} + A^{0}\left(\partial_{t}\overrightarrow {E}_{(0)}-\overrightarrow{\nabla}\wedge \overrightarrow{B}\right) \right\} + \frac{m}{\kappa}\left(2\big<\dot{x}^{0}\partial_{t}\overrightarrow{E}_{(0)}\big> \right.\nonumber\\
&&\left. \hspace*{4cm} - \big<\dot{x}^{0}\left(\overrightarrow{\nabla} \wedge\overrightarrow{B}_{(0)}\right)\big> +  \big<\overrightarrow{v}\wedge \partial_{t}\overrightarrow{B}_{(0)}\big> \right).\label{equation-33}
\end{eqnarray}

\noindent To explicitly express $\overrightarrow {E}_{(1)}$ and $\overrightarrow{B}_{(1)}$ in terms of the zeroth-order terms,  using (\ref{equation-A6}) in Appendix in order to have

\begin{equation}\label{E-1}
\overrightarrow {E}_{(1)} = \overrightarrow {E} + \frac{m}{\kappa}\left<2\dot{x}^{0}\cdot\overrightarrow {E} + \left(\partial_{t}A^{0}\cdot\overrightarrow {v} - \dot{x}^{0}\partial_{t}\cdot\overrightarrow {A}\right) + \left\{\partial^{2}_{\hspace*{0.1cm} t}A^{\mu}\cdot\overrightarrow {r} - \overrightarrow {\nabla}\left(\partial_{t}A^{\mu}\right)x^{0}\right\}\dot{x}_{\mu}\right>,
\end{equation}

\begin{equation}\label{B-1}
\overrightarrow {B}_{(1)} = \overrightarrow {B} + \frac{m}{\kappa}\left<2\dot{x}^{0}\cdot\overrightarrow {B} - \left(\overrightarrow {v}\wedge\partial_{t}\overrightarrow{A}\right) - \left\{ \overrightarrow {\nabla}\left(\partial_{t}A^{\mu}\right)\wedge\overrightarrow {r}\right\}\dot{x}_{\mu}\right>.
\end{equation}

\noindent Exploiting the last two results, we can easily find the final form on the right side of  (\ref{equation-30})-(\ref{equation-33}).

 
 

\section{$\kappa$-Generalized Lorentz force}
After having obtained above a set of four first-order equations governing the behavior of electric and magnetic fields in DSR, we focus in this part on the corresponding equation of motion describing a moving charged particle under the influence of electromagnetic field. To achieve this goal, we first differentiate the commutator involving one coordinate and one velocity with respect to the time parameter, to get \cite{S.Tanimura-1}

\begin{equation}\label{equation-34}
\left[x^{\mu}, F^{\nu}\right]=m\left[x^{\mu}, \ddot{x}^{\nu}\right]=m\frac{d}{d\tau}\left[x^{\mu}, \dot{x}^{\nu}\right]-m\left[\dot{x}^{\mu}, \dot{x}^{\nu}\right].
\end{equation}

\noindent To be able to exploit (\ref{equation-34}), we need to determine all intervening commutators. Thus, to the first order in $1/\kappa$, using (\ref{equation-13}) in order to obtain

\begin{eqnarray}\label{equation-35}
\left[x^{\mu} , F^{\nu} \right]_{(1)} &=& \left[x^{\mu} , F^{\nu}\right]_{(0)} + \left[x^{\mu} , \delta F^{\nu}\right]_{(0)} + \left[\delta x^{\mu} , F^{\nu}\right]_{(0)} \nonumber\\
                                     &=& -\frac{i\hbar}{m}\frac{\partial F^{\nu}_{(1)}}{\partial \dot{x}_{\mu}} -\frac{i\hbar}{mk}p^{0}\frac{\partial F^{\nu}_{(0)}}{\partial \dot{x}_{\mu}} + \frac{1}{k} \left[p^{0}, F^{\nu}\right]_{(0)}x^{\mu}.
\end{eqnarray}

\noindent Next, although the calculation is very affordable, for elegance, the arbitrary integration function appearing in (\ref{equation-22}) is neglected until the end and injected as another arbitrary function noted below by $G^{\nu}_{(1)}(x)$. So, by means of (\ref{equation-21}) and (\ref{equation-22}), we find

\begin{eqnarray}\label{equation-36}
\left[p^{0}, F^{\nu}\right]_{(0)} &=& mq \left[\dot{x}^{0}, \left<F^{\nu\beta}(x)\dot{x}_{\beta}\right>]_{(0)} + q^{2}[A^{0}, \left<F^{\nu\beta}(x)\dot{x}_{\beta}\right>\right]_{(0)}\nonumber\\
                       &=& mq \left < F^{\nu\beta}(x) \left[\dot{x}^{0}, \dot{x}_{\beta}\right]_{(0)} +  \left[\dot{x}^{0}, F^{\nu\beta}(x)\right]_{(0)} \dot{x}_{\beta} \right > 
                       + q^{2}F^{\nu\beta}\left[x^{\alpha}, \dot{x}_{\beta}\right]_{(0)}\partial_{\alpha}A^{0} \nonumber\\
                       &=& i\hbar q \left < \partial^{0}F^{\nu\beta}\dot{x}_{\beta} \right > -\frac{i\hbar q^{2}}{m} \left( \partial^{0}A_{\beta}F^{\nu\beta}\right),
\end{eqnarray}

\noindent which, after substitution into (\ref{equation-35}), yields

\begin{equation}\label{equation-37}
\left[x^{\mu}, F^{\nu}\right]_{(1)} = -\frac{i\hbar}{m}\left\{\frac{\partial F^{\nu}_{(1)}}{\partial \dot{x}_{\mu}} + \frac{p^{0}}{\kappa} \frac{\partial F^{\nu}_{(0)}}{\partial \dot{x}_{\mu}} + \frac{q^{2}}{\kappa} \left( \partial^{0}A_{\beta}F^{\nu\beta}\right)x^{\mu} -\frac{qm}{\kappa}\left<\partial^{0}F^{\nu\beta}\dot{x}_{\beta}\right> x^{\mu} \right\}.
\end{equation}

\noindent Furthermore, to determine the differentiated term appearing on the right side of  Eq. (\ref{equation-34}), we proceed similarly by beginning from

\begin{equation}\label{equation-38}
\left[x^{\mu}, \dot{x}^{\nu}\right]_{(1)} = \left[x^{\mu}, \dot{x}^{\nu}\right]_{(0)} + \left[x^{\mu}, \delta\dot{x}^{\nu}\right]_{(0)} + \left[\delta x^{\mu}, \dot{x}^{\nu}\right]_{(0)}.
\end{equation}

\noindent Exploiting once more Eqs. (\ref{equation-21}) and (\ref{equation-22}), Eq. (\ref{equation-20}) gives

\begin{eqnarray}\label{equation-39}
\hspace*{-0.4cm}\left[x^{\mu}, \delta\dot{x}^{\nu}\right]_{(0)} &=& \left[x^{\mu}, \frac{q}{2\kappa}\left\{\left(\left<\partial^{0}A^{\beta}\dot{x}_{\beta}\right>x^{\nu} + x^{\nu} \left<\partial^{0}A^{\beta}\dot{x}_{\beta}\right>  \right) \right.\right.\nonumber\\
&& \left.\left.\hspace*{0.7cm} + \left(A^{0}\dot{x}^{\nu} + \dot{x}^{\nu}A^{0} \right)\right\} + \frac{m}{2\kappa}\left(\dot{x}^{0}\dot{x}^{\nu} + \dot{x}^{\nu}\dot{x}^{0}\right)\right]_{(0)}\nonumber\\
&=& \frac{m}{\kappa}\left(\dot{x}^{0}\left[x^{\mu}, \dot{x}^{\nu}\right]_{(0)} + \left[x^{\mu}, \dot{x}^{0}\right]_{(0)}\dot{x}^{\nu}\right) + \frac{q}{\kappa} \left(A^{0}\left[x^{\mu}, \dot{x}^{\nu}\right]_{(0)} + \left[x^{\mu}, \dot{x}_{\beta}\right]_{(0)}\partial^{0}A^{\beta}x^{\nu}\right)\nonumber\\
&=& -\frac{i\hbar}{m\kappa} \left\{m \left(\eta^{\mu\nu}\dot{x}^{0} + \eta^{\mu 0}\dot{x}^{\nu}\right) + q\left(\eta^{\mu\nu}A^{0} + \partial^{0}A^{\mu}x^{\nu}\right)\right\}.
\end{eqnarray}
 
\noindent In the same way, Eq. (\ref{equation-13}) yields
 
\begin{equation}\label{equation-40}
\left[\delta x^{\mu}, \dot{x}^{\nu}\right]_{(0)}=-\frac{i\hbar}{m\kappa}\left\{\eta^{\mu\nu}p^{0} + q\partial^{0}A^{\nu}x^{\mu}\right\}.
\end{equation} 

\noindent Replacing Eqs. (\ref{equation-39}) and (\ref{equation-40}) into (\ref{equation-38}), we obtain

\begin{equation}\label{equation-41}
\left[x^{\mu}, \dot{x}^{\nu} \right]_{(1)} = -\frac{i\hbar}{m} \left\{\eta^{\mu\nu} + \frac{m}{\kappa}\left(2\eta^{\mu\nu}\dot{x}^{0} + \eta^{\mu 0} \dot{x}^{\nu}\right) + \frac{q}{\kappa}\left(2\eta^{\mu\nu} A^{0} + \partial^{0}A^{\mu}x^{\nu} + \partial^{0}A^{\nu}x^{\mu} \right) \right\}.
\end{equation}

\noindent Unlike Fock's nonlinear relativity where the symmetry between indices $\mu$ and $\nu$  is kept intact with respect to the deformation induced by the radius of the universe $R$ \cite{Takka-Bouda-Foughali-1} and  \cite{Takka-Bouda-2}, we note here that $\left[x^{\mu}, \dot{x}^{\nu} \right]_{(1)}$ does not in general equal $ \left[x^{\nu}, \dot{x}^{\mu} \right]_{(1)}$. It is also notable that the symmetric corrective part in (\ref{equation-41}) has a similar form as that found in \cite{Takka-Bouda-Foughali-1}. Now, taking into account Eq. (\ref{equation-22}) in such a way that the differentiation of (\ref{equation-41}) gives

\begin{eqnarray}\label{equation-42}
\frac{d}{d \tau}\left[x^{\mu}, \dot{x}^{\nu} \right]_{(1)} = &-& \frac{i\hbar q}{m\kappa} \left( \eta^{\mu 0} \left<F^{\nu\lambda}\dot{x}_{\lambda}\right> + 2\eta^{\mu\nu} \left<\partial^{0} A^{\lambda}\dot{x}_{\lambda}\right> + \eta^{\mu\lambda}\left<\partial^{0}A^{\nu}\dot{x}_{\lambda}\right>  \right. \nonumber\\  
&+& \left.  \eta^{\nu\lambda}\left<\partial^{0}A^{\mu}\dot{x}_{\lambda}\right> + \left<\partial_{\lambda} \partial^{0}A^{\mu}x^{\nu}\dot{x}^{\lambda}\right> + \left<\partial_{\lambda}\partial^{0}A^{\nu}x^{\mu}\dot{x}^{\lambda}\right>  \right).
\end{eqnarray} 

\noindent Finally, making use of relation (\ref{equation-A6}) in Appendix, the substitution of Eqs. (\ref{equation-17}) and (\ref{equation-42}) into (\ref{equation-34}) yields

\begin{equation}\label{equation-43}
F^{(1)}_{\nu}=q\big<\tilde{F}_{\nu\mu}\dot{x}^{\mu}\big> + m\big< \int\Gamma_{\nu\mu\lambda} \dot{x}^{\lambda} d\dot{x}^{\mu}\big> + G_{\nu}^{(1)}(x),
\end{equation}

\noindent where

\begin{equation}\label{equation-44}
\tilde{F}_{\nu\mu} = F_{\nu\mu} - \frac{q}{\kappa}\left(A^{0}F_{\nu\mu} + \partial^{0}A^{\lambda}F_{\nu\lambda} x_{\mu}\right)
\end{equation}
 
\noindent and
 
\begin{eqnarray}\label{equation-45}
 \Gamma_{\nu\mu\lambda} &=& \frac{q}{\kappa} \Big\{\left(\eta_{\mu 0}F_{\nu\lambda} + \eta_{\lambda 0}F_{\nu\mu}\right) + 2\left(\eta_{\nu\mu}\partial_{0}A_{\lambda} + \eta_{\nu\lambda}\partial_{0}A_{\mu} \right)
 + \partial_{0}\left(\partial_{\mu}A_{\lambda} + \partial_{\lambda}A_{\mu}\right)x_{\nu}\Big\}.
\end{eqnarray}

\noindent Here $G_{\nu}^{(1)}(x)$ is an arbitrary first-order function of position. Unlike the $\kappa$-Minkowski space-time \cite{Harikumar-Juric-Meljanac-1}, we can easily see that $\Gamma_{\nu\mu\lambda}$ is symmetric with respect to the permutation of  the two last indices $(\Gamma_{\nu\mu\lambda}=\Gamma_{\nu\lambda\mu})$. In other words, this means that the electromagnetic-type Christoffel symbols are inherently of the second kind and hence the use of all commutation relations of the $\kappa$-deformed phase space makes electromagnetic and gravitational forces more similar.

 
 

\section{Results and discussion}

In the present study, we have found an extended first approximation of Maxwell's equations (\ref{equation-30})-(\ref{equation-33}),  electric and magnetic fields (\ref{E-1}) and (\ref{B-1}), Lorentz force (\ref{equation-43}), (\ref{equation-44}) and (\ref{equation-45}) in DSR. In doing so, we have shown that the laws of electrodynamics depend on the particle mass as in the context of Fock-Lorentz transformation \cite{Takka-Bouda-Foughali-1}. This result is important because it reveals a common feature between FNLR and DSR  in spite of the fact that each one of them is based on two different concepts, the radius of the universe and Planck energy, respectively. As a relative difference, it is  notable that mass and charge intervene linearly and independently of each other in the $R$-deformed Lorentz force but as a product in DSR. Obviously, the same trend was found in $\kappa$-Minkowski space-time which incidentally constitutes the DSR space-time \cite{Harikumar-Juric-Meljanac-1}. This new effect acts so that the generalized Lorentz force is decomposed into two different contributions. The first term linear in velocity is named the electric-type Lorentz force and the second one proportional to the square of the velocity is interpreted as the gravitational-type Lorentz force. Unlike FNLR \cite{Takka-Bouda-Foughali-1} and \cite{Takka-3}, we also note here that the intervening corrective terms in the $\kappa$-generalized Lorentz force all depend on the electromagnetic field. To have a clear picture about the nature of the corrective terms appearing in our final results, let us take the most simplest case characterized by the classical static limit. In other words, we consider two identical charged particles $q$ at rest and separated by a distance $r$. Since $\dot{x}^{0}=1$ and $\dot{x}^{i}=0$, we can easily check that the use of Eqs. (\ref{equation-44}) and (\ref{equation-45}) into (\ref{equation-43}) gives

\begin{equation}\label{equation-C1}
F_{(1)}=\frac{q^{2}}{4\pi r^{2}}\left(1+\frac{2m}{\kappa}-\frac{q^{2}}{4\pi \kappa r}\right).
\end{equation}

\noindent In contrast to $\kappa$-Minkowski space-time where the $\kappa$-deformed Coulomb's law contains only one corrective term \cite{Harikumar-Juric-Meljanac-1}, we note here the existence of two terms of different types. One is proportional to the product of charge and mass and another only on charge. This suggests that two identical charged particles, submitted to the same electromagnetic field, will not feel necessarily the same force if their masses are different. Interestingly, by comparing (\ref{equation-C1}) whose construction is based on one of the last versions of DSR \cite{Ghosh-Pal-1} with that previously found in  \cite{Harikumar-Juric-Meljanac-1}, we can see a common term preceded by $-1$ and appearing as a consequence of the $\kappa$-deformed space-time. In fact, to go from one convention to another, we must multiply the commutator involving two coordinates by $-1$.  The new term depending only on charge is related to the consideration of other commutators of the $\kappa$-deformed phase space. Finally, from (\ref{E-1}) and (\ref{B-1}), we can see that $\kappa$-deformed electrostatic and magnetostatic fields are written as follows

\begin{align}
 \overrightarrow{E}_{(1)}&=\left(1+\frac{2m}{\kappa}\right) \overrightarrow{E}_{(0)}, \label{equation-C2}\\
 \overrightarrow{B}_{(1)}&=\left(1+\frac{2m}{\kappa}\right) \overrightarrow{B}_{(0)}. \label{equation-C3}
\end{align}

\noindent Surprisingly, Eqs. (\ref{equation-C2}) and (\ref{equation-C3}) share the same corrective term with (\ref{equation-C1}). If we substitute Eqs. (\ref{equation-C2}) and (\ref{equation-C3}) on the right side of (\ref{equation-30})-(\ref{equation-33}), we can easily calculate the corresponding static version for Maxwell's equations.

 
 

\section{Conclusion}
Choosing once more to investigate Feynman’s approach \cite{Dyson} in the framework of one of the extended formulations of special relativity (SR), namely doubly or deformed special relativity (DSR), we have succeeded to present here the first-order approximation of both Maxwell's equations and Lorentz force in this context. For this purpose, we have exploited the quantized $\kappa$-deformed phase space by which we have constructed the corresponding coordinate and momentum operators. At first sight, it seems that the similarity between electromagnetism and gravity is more visible in DSR than in the $\kappa$-Minkowski space-time.

To highlight more clearly the scientific contribution of our work, it is very useful to shed light on some important
points. First of all, the work presented here comes after a series of three papers devoted to the generalization of the
fundamental laws of electrodynamics in FNLR \cite{Takka-Bouda-Foughali-1}-\cite{Takka-3}. In the two last ones, we have succeeded to derive the exact generalization of Maxwell’s equations, Dirac’s magnetic monopole and Lorentz force. In the light of this progress, we intend to do the same in DSR by focusing first on the linear approximation. To move toward the higher-order
terms, the mathematical challenge is to find a better mechanism of symmetrization taking into account the
characteristics of this theory (mainly the rules of commutativity). Secondly, in comparison with some studies realized in the noncommutative framework, e.g., \cite{Carinena-Figueroa-1}, \cite{Harikumar-1}, \cite{Harikumar-Juric-Meljanac-1},
our work is mainly distinguished by the specificity of our physical context and the double aspect of our approach (a kind of unification between Feynman’s proof and the ordinary Lagrangian formulation via the use of all commutation relations of the phase space and the momentum). Thirdly, from a practical point of view, the relevance of our study is the proposition of new theoretical challenges to examine the scientific soundness of the invariance of Planck energy $E_{p}$ or Planck length $l_{p}$ in the nature. Mathematically speaking, the nonuse of any additional assumption beside the relativistic Newton’s law, making
possible the reproduction of the usual Lorentz force in special relativity, represents another technical prowess. In the quest for new generalizations even more satisfying, we plan to pursue this work, establish an extended form of the basic laws of gravity and investigate the existence of the magnetic charge in the near future. As a source of inspiration, we can take advantage of our exact studies \cite{Takka-Bouda-2}, \cite{Takka-3} as well as \cite{Harikumar-Juric-Meljanac-1}, \cite{Harikumar-Juric-Meljanac-2}, \cite{Harikumar-Kapoor}. Finally, many other analogies and more detailed analysis could also be envisaged as new perspectives, e.g., \cite{Cortese-Garcia-2}, \cite{Juric-Meljanac-Pikutic-1}, \cite{Govindarajan-Gupta-Harikumar-Meljanac-Meljanac}-\cite{Juric-Meljanac-Strajn-2}.

 
 

\appendix{}
\section*{Appendix}

\noindent To find the explicit first-order form of the electromagnetic tensor in DSR, we first use Eqs. \ref{equation-13} and \ref{equation-21} to get

\begin{eqnarray}\label{equation-A1}
\left[x^{\mu} , F^{\nu\lambda} \right]_{(1)} &=& \left[x^{\mu} + \delta x^{\mu}, F^{\nu\lambda} (x) + \delta F^{\nu\lambda}(x,\dot{x})\right]_{(1)}\nonumber\\
                                                &=&\left[x^{\mu}, \dot{x}^{\beta}\right]_{(0)}\frac{\partial \delta F^{\nu\lambda}}{\partial\dot{x}^{\beta}} + \frac{m}{\kappa}\left[\dot{x}^{0}, x^{\alpha}\right]_{(0)}\frac{\partial F^{\nu\lambda}}{\partial x^{\alpha}} x^{\mu} \nonumber\\
                                                &=&-i\hbar\left(\frac{1}{m}\frac{\partial \delta F^{\nu\lambda}}{\partial\dot{x}_{\mu}} - \frac{1}{\kappa}\partial^{0} F^{\nu\lambda} x^{\mu}\right).
\end{eqnarray}

\noindent  Furthermore, making use of Jacobi identity involving one coordinate and two velocities, we can make sure that Eq. \ref{equation-17} gives

\begin{equation}\label{equation-A2}
\left[x^{\mu} , F^{\nu\lambda} \right]_{(1)} = \frac{m^{2}}{i\hbar q}\left(\left[\dot{x}^{\lambda} , \left[x^{\mu} , \dot{x}^{\nu} \right]\right]_{(1)}- \left[\dot{x}^{\nu} , \left[x^{\mu} , \dot{x}^{\lambda} \right]\right]_{(1)}\right).
\end{equation}

\noindent Taking into account Eq. (\ref{equation-41}) and the zeroth-order form of (\ref{equation-17}), we find

\begin{eqnarray}\label{equation-A3}
\left[\dot{x}^{\lambda} , \left[x^{\mu} , \dot{x}^{\nu} \right]\right]_{(1)} 
&=& \frac{\hbar^{2}q}{m^{2}\kappa}\Big\{-\eta^{\mu 0} F^{\lambda\nu} + \left(2\eta^{\mu\nu} \partial^{0}A^{\lambda} + \eta^{\lambda\nu} \partial^{0}A^{\mu} + \eta^{\lambda\mu} \partial^{0}A^{\nu}\right)\nonumber\\
 &+& \left(\partial^{\lambda}\partial^{0} A^{\mu}x^{\nu} + \partial^{\lambda}\partial^{0} A^{\nu}x^{\mu}\right)\Big\},
\end{eqnarray}

\noindent which, after interchanging $\lambda$ with $\nu$, gives 

\begin{eqnarray}\label{equation-A4}
\left[\dot{x}^{\nu} , \left[x^{\mu} , \dot{x}^{\lambda} \right]\right]_{(1)} 
&=& \frac{\hbar^{2}q}{m^{2}\kappa}\Big\{-\eta^{\mu 0} F^{\nu\lambda} + \left(2\eta^{\mu\lambda} \partial^{0}A^{\nu} + \eta^{\nu\lambda} \partial^{0}A^{\mu} + \eta^{\nu\mu} \partial^{0}A^{\lambda}\right)\nonumber\\
 &+& \left(\partial^{\nu}\partial^{0} A^{\mu}x^{\lambda} + \partial^{\nu}\partial^{0} A^{\lambda}x^{\mu}\right)\Big\}.
\end{eqnarray}

\noindent Substituting Eqs. (\ref{equation-A3}) and (\ref{equation-A4}) into (\ref{equation-A2}), we obtain

\begin{eqnarray}\label{equation-A5}
\left[x^{\mu} , F^{\nu\lambda} \right]_{(1)} = &-& \frac{i\hbar}{\kappa}\Big\{2\eta^{\mu 0} F^{\nu\lambda} -\partial^{0}F^{\nu\lambda}x^{\mu} + \left(\eta^{\mu\nu} \partial^{0}A^{\lambda} -\eta^{\mu\lambda} \partial^{0}A^{\nu}\right)\nonumber\\  &+& \left(\partial^{\lambda}\partial^{0} A^{\mu}x^{\nu} - \partial^{\nu}\partial^{0} A^{\mu}x^{\lambda}\right) \Big\}.
\end{eqnarray}

\noindent Finally, after identification of (\ref{equation-A5}) with (\ref{equation-A1}) and integration of the resulting equation,  the generalized electromagnetic tensor takes the following form

\begin{eqnarray}\label{equation-A6}
F^{\nu\lambda}_{(1)}(x, \dot{x}) = F^{\nu\lambda}(x) &+& \frac{m}{\kappa}\left<\Big\{ 2\eta^{\mu 0}F^{\nu\lambda} + \left(\eta^{\mu\nu}\partial^{0}A^{\lambda} - \eta^{\mu\lambda}\partial^{0}A^{\nu}\right)\right.\nonumber\\
&+& \left.\left(\partial^{\lambda}\partial^{0}A^{\mu}x^{\nu} - \partial^{\nu}\partial^{0}A^{\mu}x^{\lambda}\right)\Big\}\dot{x}_{\mu} \right>.
\end{eqnarray}
 
\noindent From above, it is obvious that $F^{\nu\lambda}_{(1)}(x, \dot{x})$ is antisymmetric with respect to the permutation of indices $\nu$ and $\lambda$. Apart from the doubled corrective term, we remark also that the other terms vanish in the static case.

 
 

\end{document}